\documentclass[prd,onecolumn,nofootinbib,preprintnumbers]{revtex4}
\usepackage{amsmath}
\usepackage{amsfonts}
\usepackage{graphicx}
\usepackage{hyperref}
\usepackage{mathbbol}

\def\f{\frac}
\def\a{\alpha}
\def\be{\begin{equation}}
\def\ee{\end{equation}}
\def\bea{\begin{eqnarray}}
\def\eea{\end{eqnarray}}
\def\l{\left}
\def\r{\right}

\def\no{\nonumber}
\newcommand{\E}[1]{\times 10^{#1}}
\newcommand{\dee}{\mathrm{d}}

\renewcommand{\(}{\left(}
\renewcommand{\)}{\right)}
\renewcommand{\[}{\left[}
\renewcommand{\]}{\right]}

\begin{document}

\title{Quintessential power-law cosmology: dark energy equation of state}

\author{Burin Gumjudpai}

\affiliation{The Institute for Fundamental Study ``The Tah Poe Academia Institute"\\ Naresuan University, Phitsanulok 65000, Thailand and 
\\Thailand Center of Excellence in Physics, Ministry of Education, Bangkok 10400, Thailand}
\email{buring@nu.ac.th} 
\date{\today}
\begin{abstract}
Power-law cosmology with scale factor as power of cosmic time, $a \propto t^{\a}$, is investigated. We review and discuss value of $\a$ obtained from various types of observation. Considering dark energy dominant era in late universe from $z < 0.5$, we use observational derived results from Cosmic
Microwave Background (WMAP7), Baryon Acoustic Oscillations (BAO) and
observational Hubble data to find power exponent $\a$ and other cosmological variables.
$\a$ is found to be $0.99 \pm 0.02$ (WMAP7+BAO+$H_0$)
and $0.99 \pm 0.04$ (WMAP7). These values do not exclude possibility of acceleration at 1$\sigma$ hence giving viability to power-law cosmology in general.
When considering scenario of canonical
scalar field dark energy with power-law cosmology, we derive scalar field potential, exact scalar field solution and equation of state parameter. We found that the scenario of power-law cosmology containing dynamical canonical scalar field predicts present equation of state parameter $w_{\phi, 0} =  -0.449 \pm 0.030 $
while the $w$CDM with WMAP7 data (model independent, $w$ constant) allows a maximum (+1$\sigma$) value of $w_{\phi, 0}$ at -0.70 which is off the prediction range. However, in case of varying $w_{\phi}$, the $w_{\phi, 0}$ value predicted from quintessential power-law cosmology is allowed within 1$\sigma$ uncertainty.    
\end{abstract}

\maketitle

\section{Introduction}

In physics, scalar field matter plays a key role in symmetry-breaking mechanisms while in cosmology it  contributes to acceleration expansion of space. In the early universe, scalar field dynamics drives super-fast expansion in inflationary scenario, resolving horizon and flatness problems as well as explaining the origin of structures 
\cite{inflation} \cite{inflation2} \cite{inflation3} \cite{inflation4} \cite{inflation5}.  The scalar field is also believed to be responsible for present acceleration in various models of dark energy \cite{Padmanabhan:2004av} \cite{Copeland}.  The present acceleration has been observed  by various observations, e.g. the cosmic microwave background (CMB) \cite{Masi:2002hp} \cite{Larson:2010gs} \cite{arXiv:1001.4538}, large-scale structure surveys \cite{Scranton:2003in} \cite{Tegmark}, supernovae type Ia  (SNIa) \cite{Perlmutter:1997zf} \cite{Perlmutter:1999} \cite{Riess:1998cb} \cite{gold} \cite{tonry} \cite{RiessGold2004} \cite{Astier:2005qq} \cite{Amanullah2010} and X-ray luminosity from galaxy clusters \cite{Allen:2004cd} \cite{Rapetti}. Simplest way to explain the present acceleration is to introduce a cosmological constant into the field equation \cite{cosmoconstant} \cite{cosmoconstant2} \cite{cosmoconstant3}, but the idea suffers from the fine-tuning problem \cite{SahniSta2000} \cite{Peebles}. In order for the cosmological constant to be viable, there is a proposed model of varying cosmological constant \cite{varcc} \cite{varcc2} which is not necessary a part of cosmological constant paradigm.

Here we consider scenario scale factor is assumed to be 
function of the cosmic time as $a \propto t^{\a},\; 0 \leq \a  \leq \infty $ at late universe from $z < 0.5$.   
Fundamental motivations of the power-law cosmology are such as non-minimally coupled scalar-tensor theory in which the scalar field couples to the curvature contributing to energy density that cancels out the vacuum energy \cite{Dolgov1982} \cite{Fujii1990}  and in simple inflationary model in which the power-law can remove flatness and horizon problems with simple spectrum \cite{Lucchin}.  In linear-coasting case, $ \a  \approx 1$  \cite{Lohiya:1996wr} \cite{Sethi:1999sq} \cite{Dev:2000du} \cite{Dev:2002sz}, fundamental motivations come from SU(2) instanton cosmology \cite{Allen:1998vx}, higher order (Weyl)  gravity \cite{Manheim}, or from scalar-tensor theories \cite{Lohiya:1998tg}. However, in the early universe the $\a$ value is tightly constrained  by big bang primordial nucleosynthesis (BBN). In order to be capable of light element abundances, maximum $\a$ allowed is approximately $0.55$ (see Refs. \cite{Kaplinghat:1998wc} and \cite{Kaplinghat:2000zt}). This value results in much younger cosmic age and clearly does not give acceleration.  The large $\a$ model was proposed long ago by Kolb  \cite{Kolb1989} to resolve age problem in CDM model. It also evades the flatness and horizon problems. Hence if the power-law cosmology is to be valid, the exponent $\a$ should not be constant but evolving.  The power-law expansion is often used in astrophysical observations since its convenience is in adjustability of the rate of expansion characterized by only one parameter, $\alpha$. Therefore power-law model is a good phenomenological description of the cosmic evolution \cite{Peebles:1994xt} as it can describe radiation epoch, dark matter epoch, and dark energy epoch according to value of the exponent. At each epoch, different matter-energy contents lead to different power-law exponents.  
We know that $\a \approx {1/2}$ in radiation era and after the matter-radiation equality era, $z \lesssim 3196$ (value from Ref. \cite{Larson:2010gs}), one can relax the BBN constraint and the universe evolves with $\a \approx {2/3}$. Until recent past when dark energy began to dominate $z \lesssim 0.5$, $\a \gtrsim 1$ is required so that it can give late acceleration. 

Power-law cosmology is also considered in specific gravity or dark energy models such as in $f(T)$ and $f(G)$ gravities \cite{SetareAstrophysicsSpace} \cite{SetareAstrophysicsSpace2} and in the case of which there is coupling between cosmic fluids \cite{arXiv:0803.1086}. The power-law cosmology were also studied in context of scalar field cosmology \cite{GumjudpaiPowerlaw1}, phantom scalar field cosmology \cite{arXiv:1008.2182}. There is also slightly different form of the power-law function which $\a$ can also evolved with time so that it can parameterize cosmological observables \cite{astro-ph/0405368}. Considering power-law expansion in these models is useful for estimating the other unknown parameters such as coupling constant of the gravity models and it is reasonable when the $\a$ is approximately taken as an average value over a short range of redshift.

Here we investigate scenario similar to an analysis previously done for closed geometry \cite{GumjudpaiPowerlaw1} in which derived results from WMAP5 and WMAP5+BAO+SNIa combined datasets are used.   Here we consider flat universe and we use results from WMAP7 (Ref. \cite{Larson:2010gs}) and WMAP7+BAO+$H_0$ combined datasets (Ref. \cite{arXiv:1001.4538}) in order to constrain equation of state parameter of the scalar-field power-law cosmology. The WMAP7 results are presented in Table \ref{datatable}. Due to large systematic error of the SNIa data, i.e. comparable to statistical error, the SNIa data is not incorporated in the WMAP7 data \cite{Larson:2010gs}. This is good and bad at the same time. Without using SNIa data in CMB combined analysis, one can not constrain curvature value and the flat case is hence assumed.   We present other data such as critical and dust matter densities in section \ref{Sec_W7} and determine value of $\a$  in section \ref{Sec_Power}. We give summary and comments on the value of  $\a$ found in previous literatures. We consider scalar-field power-law cosmology in which canonical (quintessence) scalar field evolving under potential $V(\phi)$ and dust barotropic fluid (cold dark matter and baryonic matter) are two major ingredients in section \ref{Sec_field}. Considering time after dark energy domination, i.e. $z \lesssim 0.5$ with approximately constant power-law exponent, we determine scalar field equation of state parameter, $w_{\phi}$ at present in section \ref{Sec_resultP}. The plots of results from WMAP7, WMAP7+BAO+$H_0$ and WMAP5+BAO+SNIa (previous work with closed geometry case) are presented for comparison.  At last we conclude this work in section \ref{Sec_con}.

 \section{Cosmological parameters}  \label{Sec_W7}
Cosmological parameters are fixed by values at present (subscripted with 0) and we set $a_0=1$ here.
The values of $H_0$, $t_0$,  $\Omega_{\mathrm{CDM}, 0}$, and $\Omega_{b, 0}$ are of derived data obtained from WMAP7 data  \cite{Larson:2010gs} and WMAP7 combined data with Baryon Acoustic Oscillations (BAO) and $H_0$ data
\cite{arXiv:1001.4538} of which we take the maximum likelihood value assuming spatially flat case. Although in deriving the value of $t_0$, the $\Lambda$CDM model is assumed when exploiting the CMB data, it can be estimably used since present $w$ of dark energy is very close to -1.   Total present dust density parameter is summing of baryonic and cold dark matter components that is $
 \Omega_{\rm m, 0} = \Omega_{\mathrm{CDM}, 0} + \Omega_{b, 0}$.
We define
$ D \equiv \rho_{\rm m, 0} = \Omega_{\rm m, 0} \rho_{c, 0}\,.  $ and  $\rho_{c, 0} \equiv 3H_0^2/8\pi G$ is  present value of the critical density. Radiation and other neutrino densities are negligible here. These are presented in Table \ref{datatable}.

\begin{table*}[!h]
\begin{ruledtabular}
\begin{tabular}{ccc}
Parameter & WMAP7+BAO+$H_0$ & WMAP7  \\
\hline
$t_0$ & $13.76\pm0.11$ Gyr or $(4.34\pm0.03) \times 10^{17}$ sec & $13.79\pm0.13$  Gyr  or $(4.35\pm0.04) \times 10^{17}$ sec \\
$H_0$ & $70.4\pm1.4$ km/s/Mpc & $70.3\pm2.5$ km/s/Mpc\\
$\Omega_{b,0}$ & $0.0455\pm0.0016$ & $0.0451\pm0.0028$\\
$\Omega_{\mathrm{CDM},0}$ & $0.226\pm0.015$ & $0.226\pm0.027$\\
\hline
$\rho_{\rm m, 0}$ & $(2.53\pm0.17) \times 10^{-27}$ ${\rm kg/m^3}$ & $(2.52\pm0.31) \times 10^{-27}$ ${\rm kg/m^3}$\\
    $\rho_{c, 0}$ & $(9.31\pm0.37)\times 10^{-27}$ ${\rm kg/m^3}$ & $(9.28\pm0.66) \times 10^{-27}$  
     ${\rm kg/m^3} $ \\
\end{tabular}    
\end{ruledtabular} \label{datatable}
\caption{Combined WMAP7+BAO+$H_0$ and WMAP7 derived parameters from Refs. \cite{Larson:2010gs} and \cite{arXiv:1001.4538}. Present dust density and present critical density obtained from WMAP7 data are also shown here. }
\end{table*}

\section{Power-Law Cosmology}  \label{Sec_Power}
In power-law cosmology, scale factor is a function of time as
\begin{equation}\label{scalefactor}
 a(t) = a_0 \left( \frac{t}{t_0} \right)^{\a}\,, \end{equation}
The Hubble parameter is $ H(t) = \dot{a}/{a} = {\a}/{t} $ with
acceleration $ \dot{H} = -\a/t^2 $. Using fixed value at present,  $\a$  is simply $H_0 t_0$.
The deceleration parameter in this scenario is
\begin{equation}\label{decel}
q\: \equiv \: - \frac{a \ddot{a}}{\dot{a}^2}\: =\:  \f{1}{\a} -1 \,,
\end{equation}
that is $\a = 1/(q+1)$. As $\a \geq 0$ is required in power-law cosmology, hence $q \geq -1$ and $H_0 \geq 0$.
There have been attempts to indicate the value of $\a$.  Typically astrophysical tests for the power-law cosmology can be performed using
gravitational lensing statistics \cite{Dev:2002sz}, high-redshift objects such as distant globular clusters, SNIa
\cite{Sethi2005} \cite{Dev:2008ey} \cite{Kumar2011}, compact-radio source \cite{Jain2003} or using X-ray gas mass fraction
measurements of galaxy clusters \cite{Allen2002} \cite{Allen2003} \cite{Zhu:2007tm}. Study of angular size to $z$ relation of a large sample of milliarcsecond compact radio sources in flat FLRW universe found that $\a =  1.0 \pm 0.3$ at 68 \% C.L. \cite{Jain2003}. X-ray mass fraction data of galaxy clusters for flat power-law cosmology gives $\a = 2.3^{+1.4}_{-0.7} $ (Ref. \cite{Zhu:2007tm}) and a joint test using Supernova Legacy Survey (SNLS) and $H(z)$ data in flat case gives $\a = 1.62^{+0.10}_{-0.09}$ (Ref. \cite{Dev:2008ey}). WMAP5 dataset gives $\a =  1.01$ (closed geometry) \cite{GumjudpaiPowerlaw1}. Some of these values of  $\a$ are found under specific assumption of spatial curvature. We summarize this in Table \ref{tableA}. When data is spatial-curvature independent, the geometry type is not specified in the table.

\begin{table*}[!h]
\begin{ruledtabular}
\begin{tabular}{cccccc}  
Refs.                                           &  Data             &    $q$                                           &   $H_0$  {\footnotesize(km/sec/Mpc)}                   & $\alpha$         &  $t_0$       \\
\hline
 \footnotesize{\cite{Sethi2005}} &    SNIa \footnotesize{(Gold Sample)}       &  $ -$             &   $-$      &   $ 1.04^{+0.07}_{-0.06}$ {\footnotesize{(open)}}       &  $-$     \\
      & quasar age estm.\footnotesize{(APM 08279+5255)}  &  $ -$             &   $-$      &   $ \geq 0.85   $      &  $-$     \\ \hline
 &  cluster gas mass frac.\footnotesize{(Chandra)}   &  $ -$             &   $-$      &   $ 1.14^{+0.05}_{-0.05}$ \footnotesize{(open)} &  $-$ \\
   \footnotesize{ \cite{Zhu:2007tm}} &  cluster gas mass frac.\footnotesize{(Chandra)}   &  $ -$  &   $-$  &   $ 2.3^{+1.4}_{-0.7}$ \footnotesize{(flat)}  &  $-$ \\
 &  cluster gas mass frac.\footnotesize{(Chandra)}   &  $ -$  &   $-$  &   $ 0.95^{+0.06}_{-0.06}$ \footnotesize{(closed)}  &  $-$ \\  \hline
                              &  SNIa \footnotesize{(SNLS)}   &  $ -$             &   $-$      &   $    1.42^{+0.08}_{-0.07} $ \footnotesize{(open)}        &  $-$     \\
             &  $H(z)$ \footnotesize{(GDDS+archival)}   &  $ -$             &   $-$      &   $ 1.07^{+0.11}_{-0.09}   $      &  $-$     \\
     \footnotesize{ \cite{Dev:2008ey}}         & $H(z)+$SNIa    &  $ -$             &   $-$      &   $   1.31^{+0.06}_{-0.05} $ {\footnotesize{(open)}}      &  $-$     \\
                     & $H(z)+$SNIa    &  $ -$             &   $-$      &   $   1.62^{+0.10}_{-0.09} $ {\footnotesize{(flat)}}      &  $-$     \\
                         &  $H(z)+$SNIa    &  $ -$             &   $-$      &   $  2.28^{+0.23}_{-0.19} $ {\footnotesize{(closed)}}      &  $-$  
                            \\ \hline
 \cite{GumjudpaiPowerlaw1}                        &    WMAP5   &  $ -$             &   72.4      &   $    1.01 $ \footnotesize{(closed)}        &  13.69 Gyr     \\   
 &  WMAP5+BAO+SNIa &  $ -$  & 70.2  &  $  0.985   $  \footnotesize{(closed)}  & 13.72 Gyr  \\ \hline
    &   $H(z)$  {\footnotesize{(new GDDS+archival)}}   &   $-0.10^{+0.13}_{-0.14}  $          &   $65.18^{+3.12}_{-2.98} $  &  $ 1.11^{+0.21}_{-0.14}  $     &  $  16.65^{+3.25}_{-2.23}$  Gyr    \\
 \footnotesize{ \cite{Kumar2011}}  &  SNIa {\footnotesize{(Union2)} }                     &      $  -0.38^{+0.05}_{-0.05}  $        &   $69.18^{+0.55}_{-0.54} $     &  $    1.61^{+0.14}_{-0.12} $  \footnotesize{(flat)}       &  $22.76^{+1.99}_{-1.71}$  Gyr     \\
  &   $H(z)+$SNIa            &        $-0.34^{+0.05}_{-0.05}     $   &   $68.88^{+0.53}_{-0.52}$     &   $    1.52^{+0.12}_{-0.11} $  \footnotesize{(flat)}   &  $ 21.58^{+1.71}_{-1.57} $ Gyr   \\ \colrule
\footnotesize{This article}                        &    WMAP7   &  $ -$             &   $70.3^{+2.5}_{-2.5}$     &   $ 0.99^{+0.04}_{-0.04}  $ \footnotesize{(flat)}        &  $13.79^{+0.13}_{-0.13}$ Gyr     \\
     &  WMAP7+BAO+$H(z)$ &  $ -$  & $70.4^{+1.4}_{-1.4}$  &  $  0.99^{+0.02}_{-0.02}   $  \footnotesize{(flat)}  &  $13.76^{+0.11}_{-0.11}$ Gyr          \\
  \end{tabular}    
   \label{tableA}
\caption{Obervational data constraint for power-law cosmology:  All analysis for $q, H_0, \a, t_0$ are spatial-curvature independent except for studies of SNIa data and of cluster X-ray gas mass fraction. In deriving the SNIa luminosity distance relation to redshift, the result depends on curvature assumption. In Ref. \cite{Kumar2011}, $H(z)$ analysis uses 15 data points (from Ref. \cite{Cao2011}) and 557 SNIa data points (from Union2 dataset of Supernova Cosmology Project in Ref. \cite{Amanullah2010}) were used with flat spatial curvature assumption. 
In completion of Ref. \cite{Kumar2011}, we present $\a$ and $t_0$ here.
In  Ref. \cite{Dev:2008ey}, $H(z)$ values are from  Ref. \cite{ref:SVJ2005} which took 32 data points from the Germini Deep Deep Survey (GDDS) and archival data to obtain 9 data points at  $0.09 \leq z\leq 1.75 $.
Studies of power-law cosmology with SNIa data (from Gold Sample \cite{Sethi2005}
 (157 data points from Ref. \cite{RiessGold2004}), Supernova Legacy Survey  \cite{Dev:2008ey} (SNLS, 115 data points from Ref. \cite{Astier:2005qq})) and X-ray gas mass fractions in galaxy clusters   \cite{Zhu:2007tm} show that the open model is favored, flat and closed models are not ruled out.  Without power-law cosmology assumption and without pre-assumed geometry, WMAP5 data suggests that the universe is slightly closed  \cite{GumjudpaiPowerlaw1} \cite{Komatsu:2008hk}.}
 \end{ruledtabular} 
\end{table*}

We should notice that when $\a$ is found with curvature-independent procedure
(i.e. with neither SNIa nor cluster X-ray gass mass fraction) or in flat case, $\a$ value is very near unity. For example, $H(z)$ data gives $\a = 1.07^{+0.11}_{-0.09}$ (Ref. \cite{Dev:2008ey}) and  $\a = 1.11^{+0.21}_{-0.14}$ (Ref. \cite{Kumar2011}). For the flat case, WMAP7 gives $\a = 0.99^{+0.04}_{-0.04}$ and WMAP7 combined result gives  $\a = 0.99^{+0.02}_{-0.02}$.  Inclusion of SNIa data in combined analysis would render greater value of $\a$ (see in the Table \ref{tableA}). Although, investigation of power-law cosmology model with SNIa data  (in Refs. \cite{Sethi2005} and  \cite{Dev:2008ey}) and with X-ray gas mass fractions in galaxy clusters (in Ref. \cite{Zhu:2007tm}) favor  open power-law cosmology model but flat and closed cases are still not ruled out.  It would be an improvement if chi-square parameter of larger number of SNIa data points (e.g. Union2) are analyzed with $H(z)$ data for open and closed cases as done for flat case in Ref. \cite{Kumar2011}. Then one can tell more precisely whether the open power-law cosmology is favored over the flat and closed ones.  Larger SNIa data points in combined analyzed with latest WMAP dataset would distinct the cosmic geometry.

\section{Scalar-Field Power-Law Cosmology} \label{Sec_field}
In this section we consider CDM model with zero cosmological constant of the late FLRW universe. Two fluid components, cold dark matter and homogenous canonical scalar field $\phi \equiv \phi(t)$ are ingredients of the universe. Dynamics of the barotropic fluid is governed by the fluid equation $ \dot{\rho}_{\rm m} =
-3H\rho_{\rm m}, $ and
\begin{equation}\label{energydensity} \rho_{\rm m} = \frac{D}{a^n}, \end{equation}
for a constant $n \equiv 3 (1 + w_{\rm m})$.  $D \geq 0$ is a proportional constant. The scalar field  is minimally coupled to
gravity with Lagrangian density
$ {\mathcal{L}}_{\phi} = - ({1}/{2})\partial_{\mu} \phi \, \partial^{\mu} \phi   -  V(\phi)\,. $
 The field action, $ S_{\phi} = \int {\rm d}^4 x \, {{\mathcal{L}}}_{\phi}$, with variation $\delta S = 0$ gives  field equation of motion
\begin{equation}
 \ddot{\phi} + 3H\dot{\phi} + \frac{\dee}{\dee \phi}V = 0.
\end{equation}
describing energy conservation of the field as the universe is expanding. Here
scalar field energy density and scalar field pressure
\begin{align}
\rho_\phi = \frac{1}{2}  \dot{\phi}^2 + V(\phi),\quad  p_\phi = \frac{1}{2}  \dot{\phi}^2 - V(\phi).   \label{rhophipphi}
\end{align}
Total density  and total pressure are just addition of the density or pressure of the two components.
The Friedmann equation is just
\begin{equation}
H^2 = \frac{8 \pi G}{3} \rho_\mathrm{tot}  - \f{k}{a^2}
\end{equation}
The Friedmann equation can be rearranged to
\begin{equation}
\rho_{\phi} = \frac{3}{8 \pi G}\l( H^2  -  \f{8 \pi G}{3}\f{D}{a^n}  + \f{k}{a^2}  \r)  \label{laboo}
\end{equation}
The acceleration equation of this system is
\begin{equation}
\dot{H} \,=\, \f{\ddot{a}}{a} - \f{\dot{a}^2}{a^2} \,=\, - 4\pi G \l( \rho_{\rm m} + p_{\rm m}  + \rho_{\phi}  + p_{\phi}  \r) \,+\,  \f{k}{a^2}   \label{acceleration}
\end{equation}
Using (\ref{rhophipphi}) in (\ref{acceleration}) we rearrange the equation to get
\be
\dot{\phi}^2 \,=\, -\f{1}{4\pi G} \l( \dot{H} - \f{k}{a^2} \r)  - \f{n}{3} \f{D}{a^n}   \label{labee}
\ee
We insert (\ref{laboo}) and (\ref{labee}) into $\rho_\phi = ({1}/{2}) \dot{\phi}^2 + V(\phi)$, it is straightforward to obtain the scalar field potential
\begin{equation}
\label{vraw}
V(\phi) = \frac{3}{8 \pi G} \left( H^2 + \frac{\dot{H}}{3}  \right) + \left( \frac{n - 6}{6} \right) \frac{D}{a^n},
\end{equation}
where $8\pi G = M_\mathrm{P}^{-2}$ and  $M_\mathrm{P}$ is the reduced Planck mass.  We consider only the flat case of which $k=0$ and the barotropic fluid is dust ($n=3$) in this work.

\section{Results} \label{Sec_resultP}
\begin{figure}[t]
\centering
\includegraphics[width=3.5in]{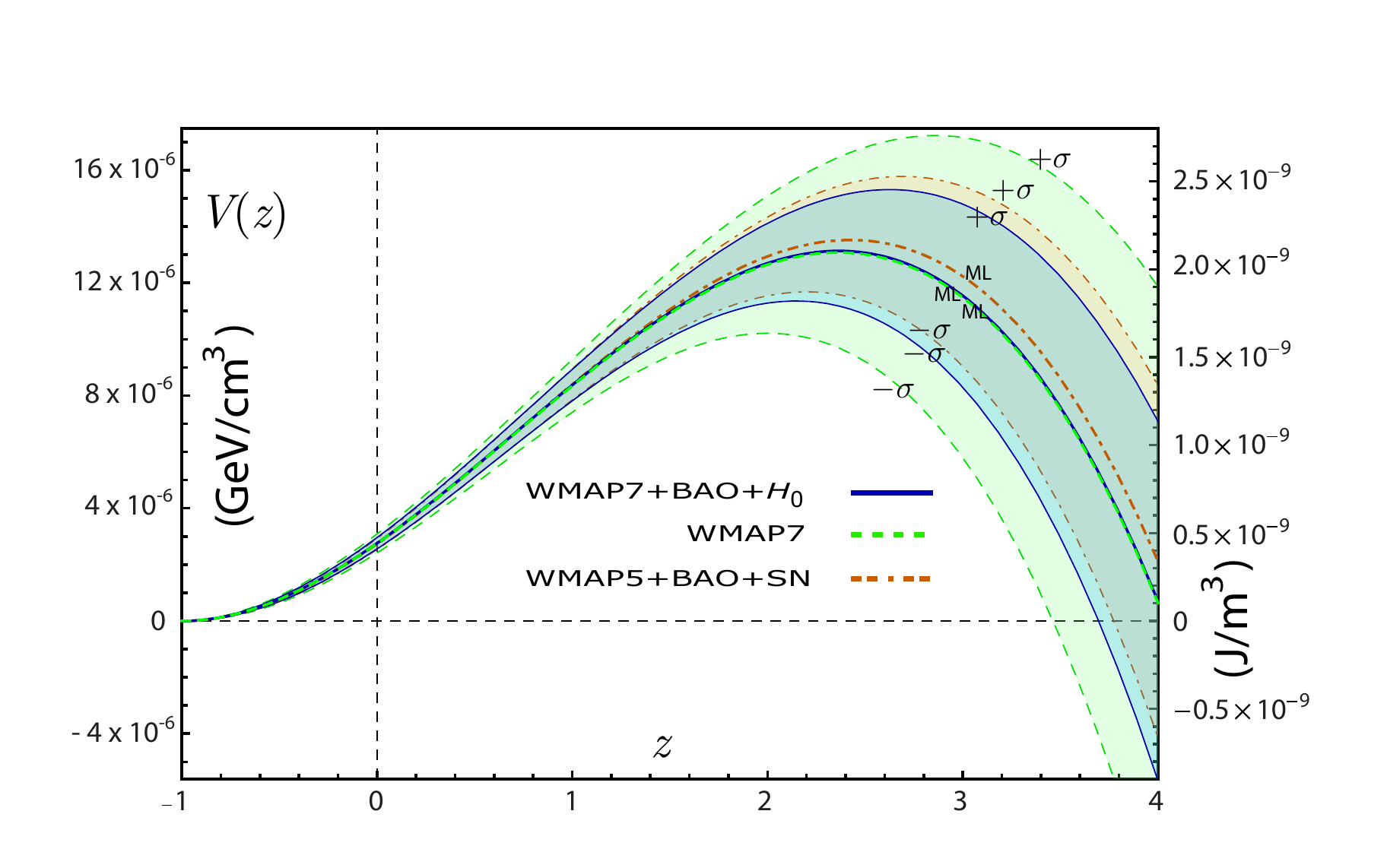}
\caption{Scalar potential plotted versus $z$ using data from three
 datasets, WMAP7+BAO+$H_0$, WMAP7 and WMAP5+BAO+SNIa and their error bar (1$\sigma$) regions}\label{fig_vz}   \end{figure}

\begin{figure}[t]
\centering
\includegraphics[width=3.5in]{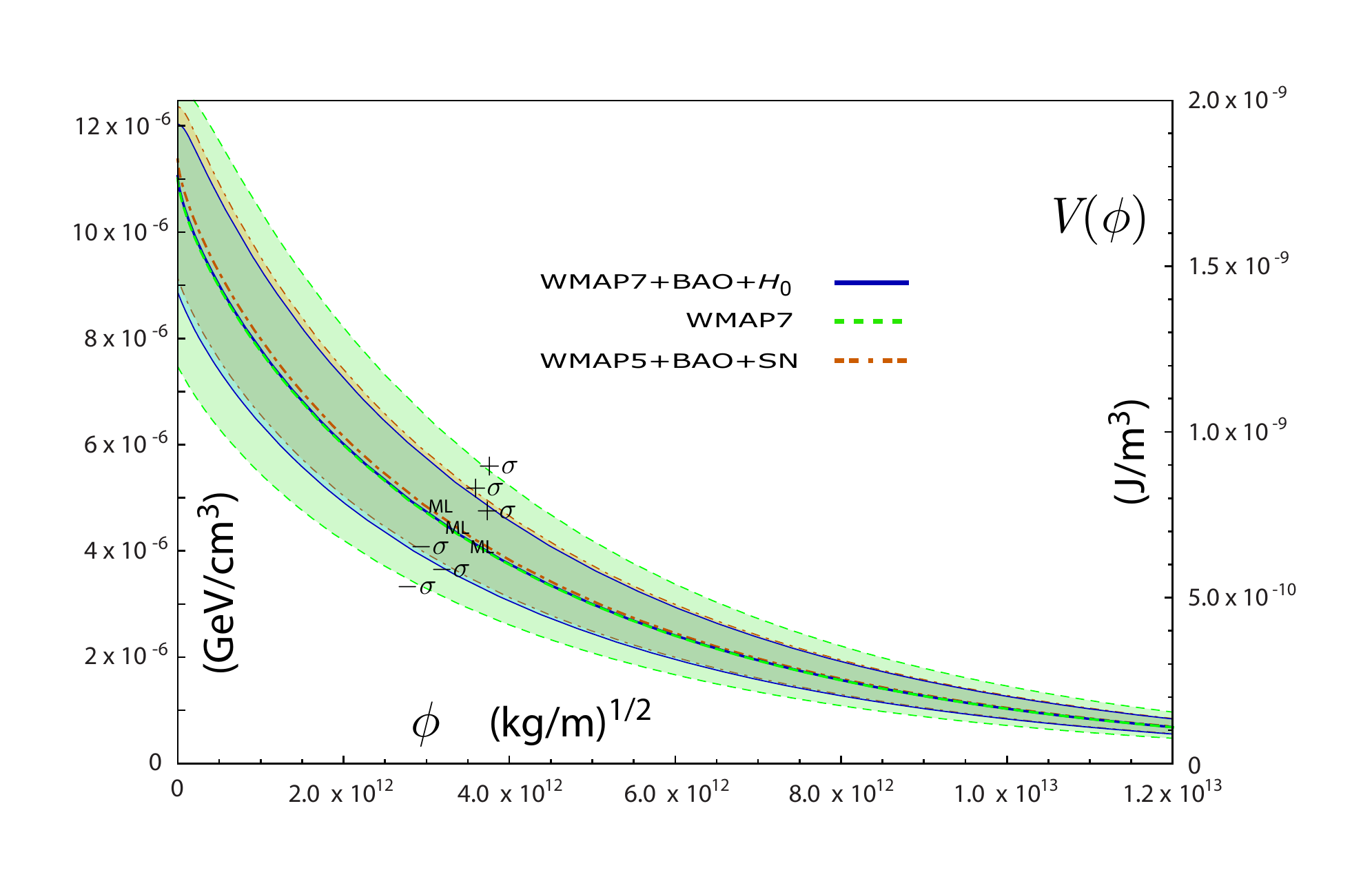}
\caption{Scalar potential plotted versus $\phi$ using data from three
 datasets, WMAP7+BAO+$H_0$, WMAP7 and WMAP5+BAO+SNIa and their error bar (1$\sigma$) regions}\label{fig_vphi}
 \end{figure}

Assuming power-law expansion, in order to find equation of state, the potential can be written down. Note that 
constructions of model-independent scalar potential were performed before by many authors for instance, developing formalism for constructing potential of a non-minimally coupled scalar field and finding equation of state using relation of distance measurement and redshifts \cite{ref:SVJ2005} \cite{Starobinsky1998} \cite{S1} \cite{S3} \cite{Saini:1999ba} \cite{Chiba} \cite{Alam:2003fg} \cite{astro-ph/0506696}  \cite{Li:2006ea}. Other potential construction are studied in different situations, such as 
the case when assuming of barotropic density as scaling function of scale facfor \cite{Liddle:1998xm} \cite{Rub:2001}, non-flat universe potential construction from late-time attractors \cite{Copeland2009}.  We do not construct scalar potential in similar manner to these references but we only use WMAP7 data to fix a present value for scalar potential considering the expansion is approximately power-law in very recent past, i.e. $z < 0.5$. In SI units, 
 $M_\mathrm{P}^2 =\hbar c / 8\pi G$, consider dust
 matter domination ($n = 3$), we write 
\begin{equation}
 V(t)\, =\, \frac{M_\mathrm{P}^2 c}{\hbar}
 \left(
 \frac{3 \a^2 - \a}{t^2}   \right)
\,  -\, \frac{Dc^2}{2}\l(\frac{t_0}{t} \r)^{3\a}\,.
 \label{vt}
 \end{equation}
Using both datasets in the tables, in power-law cosmology scenario, the scalar potential function is
(for WMAP7$+$BAO$+H_0$),
\bea
V(t) \;& = &\; \f{1.05\E{26}}{(t \;{\rm in\; sec})^2}  - \f{2.96\E{42}}{(t \;{\rm in\; sec})^{2.97}} \; \; \; {\rm J/m}^3  
\; = \;  \f{1.05\E{59}}{(t \;{\rm in\; Gyr})^2}  - \f{2.98\E{91}}{(t \;{\rm in \;Gyr})^{2.97}} \;\; \; {\rm J/m}^3  \no \\
& = &    \f{6.55\E{29}}{(t \;{\rm in\; sec})^2}     - \f{1.85\E{46}}{(t \;{\rm in\; sec})^{2.97}} \;  \;\;  {\rm GeV/cm}^3 \; =  \;  \f{6.55\E{62}}{(t \;{\rm in\; Gyr})^2} - \f{1.86\E{95}}{(t \;{\rm in \;Gyr})^{2.97}}            \;      {\rm GeV/cm}^3    \no \\
\eea
and for WMAP7,
\bea
V(t) & = & \f{1.05\E{26}}{(t \;{\rm in\; sec})^2}  -\f{3.25\E{42}}{(t \;{\rm in\; sec})^{2.97}}   \; \; {\rm J/m}^3  \; =  \;  \f{1.05\E{59}}{(t \;{\rm in\; Gyr})^2}  - \f{3.27\E{91}}{(t \;{\rm in \;Gyr})^{2.97}} \;\; \; {\rm J/m}^3  \no \\
& = &    \f{6.55\E{29}}{(t \;{\rm in\; sec})^2}     - \f{2.03\E{46}}{(t \;{\rm in\; sec})^{2.97}} \; \;\;  {\rm GeV/cm}^3
  \; =  \;  \f{6.55\E{62}}{(t \;{\rm in\; Gyr})^2} - \f{2.04\E{95}}{(t \;{\rm in \;Gyr})^{2.97}}            \;      {\rm GeV/cm}^3  \no \\
\eea
We plot potential versus redshift in Fig. \ref{fig_vz},  using conversions, $a = (1+z)^{-1}$ and $ t = t_0 (1+z)^{-1/ \a}$. Although we consider late universe at $z<0.5$ (i.e. $t \approx 9.14$ Gyr (WMAP7 combined) and $t \approx 9.16$ Gyr (WMAP7)), in our plot we show also earlier time portion for completion. 
From (\ref{labee}), the scalar field kinetic term for power-law cosmology reads
\be
{\dot{\phi}}^2\,  =\,   \f{2 M_{\rm P}^2 c }{\hbar}    \f{\a}{t^2} \,-\, {D c^2}  \l(\f{t_0}{t}\r)^{3\a}\,.
\ee
We integrate this equation to obtain scalar field solution,
\begin{eqnarray}
\phi (t) &=& -\frac{2}{3\alpha - 2}\sqrt{\frac{2M_{\rm P}^2c}{\hbar}\alpha - Dc^2t_0^{3\alpha}\left(\frac{1}{t}\right)^{3\alpha-2}}   \no \\
 & &+\: \frac{2}{3\alpha - 2}\sqrt{\frac{2M_{\rm P}^2c}{\hbar}\alpha}\,\tanh^{-1}\left[\sqrt{1-\frac{\hbar c D t_0^{3\alpha}}{2M_{\rm P}^2\alpha}\left(\frac{1}{t}\right)^{3\alpha-2}}\right]\,, 
\end{eqnarray}
%
%
to which we can use  WMAP7 and combined WMAP7 data to numerically plot $V(\phi)$ in Fig. \ref{fig_vphi}. 

\begin{figure}[t]
\centering
\includegraphics[width=3.2in]{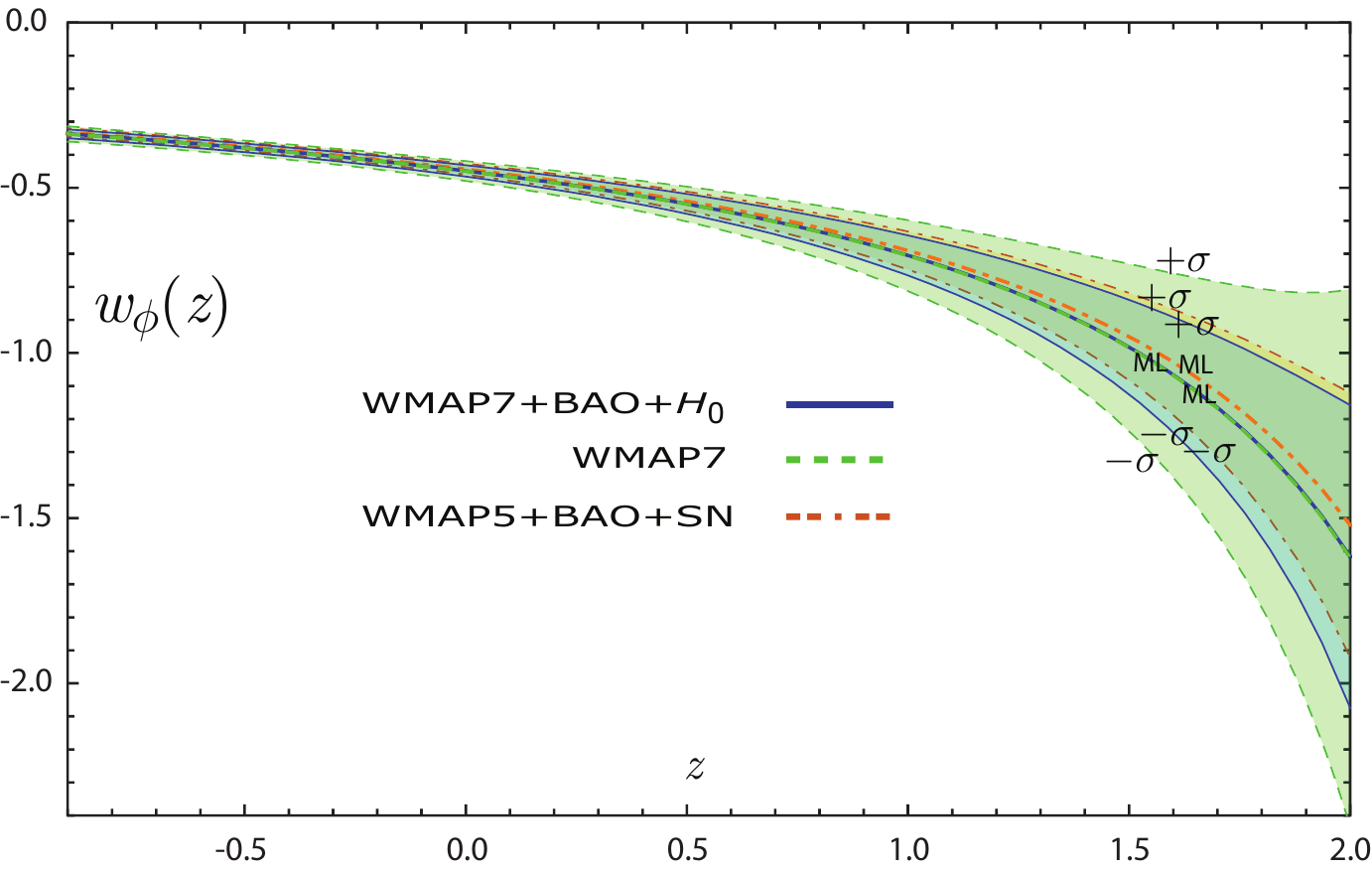}
 \caption{At present ($z = 0$), the equation of state parameter of the scalar-field power-law cosmology does not include the observation favored value ($w \approx -1$) even at 1$\sigma$ regions.}\label{w}
 \end{figure}

The equation of state parameter is found directly from
$w_{\phi} = p_{\phi}/\rho_{\phi}$ and using expression for $\dot{\phi}^2$ and $V(\phi)$ to get
\be
w_{\phi}(t) = \f{  \l( M_{\rm P}^2 c/ \hbar \r) \l[   (-3 \a^2 + 2 \a)/ t^2  \r] }{  \l( M_{\rm P}^2 c/ \hbar \r)
  \l( 3\a^2/t^2 \r)  -  Dc^2 \l(  t_0/ t  \r)^{3\a}   } \,.
\ee
We then have
\begin{equation}
w_\phi(z) = -1 + \frac{2\alpha + f(z)}{3\alpha^2+f(z)}\,,
\end{equation}
where $f(z)\equiv-(\hbar c/M_{\rm P}^2) Dt_0^2(1+z)^{(3\alpha-2)/\alpha}$.  It is found that (WMAP7$+$BAO$+ H_0$)
\bea
w_{\phi}(z) &=& \frac{1}{-3.058+0.830 (1+z)^{0.981}}\,,  \hspace{0.2cm}    \\
 w_{\phi}(z=0) &=& -0.4489\pm0.0172  \,.    \eea
and (WMAP7) 
\bea
w_{\phi}(z) &=& \f{1}{-3.053+0.828 (1+z)^{0.983}}\,,  \hspace{1.5cm}
  \\
w_{\phi}(z=0) &= & -0.4493\pm0.0300\,.
\eea
Recent evolutions of the equation of state using two dataset predicted by power-law cosmology are shown in Fig. \ref{w}. 
Note that these values of equation of state parameters are not the CMB derived value of the $w$CDM model ($w=w(a)$).
Our $w_{\phi}$ values are found in context of scalar-field power-law cosmology and these are much greater than observational (spatially flat) WMAP model-independent derived results 
which are
$w_{\phi, 0} = -1.12^{+0.42}_{-0.43}$  (WMAP7 data with constant $w$) and $w_{\phi, 0} = -1.10^{+0.14}_{-0.14} $ (68 \% CL) (WMAP7+BAO+$H_0$ with constant $w$). The other values (derived with time varying $w$) are given by WMAP7+BAO+$H_0$+SN:  $w_{\phi, 0} = -1.34^{+1.74}_{-0.36}$ and  WMAP7+BAO+$H_0$+SN with time delay distance information:  $w_{\phi, 0} = -1.31^{+1.67}_{-0.38}$. Large positive error bar is a result of large systematic error in SN data \cite{Larson:2010gs} \cite{arXiv:1001.4538}.

\section{Conclusion} \label{Sec_con}

We study power-law cosmology at late time from $z \approx 0.5$  to present. The power exponent $\a$
is approximately constant during this period. Finding $\a$  is important task in power-law cosmology as it is major feature for solving flatness, horizon and age problems in cosmology.   This is to see if it could agree with the present acceleration. Using cosmic microwave background derived maximum-likelihood cosmological parameters from WMAP7 datasets and WMAP7+Baryon Acoustic Oscillation (BAO)+$H_0$ combined dataset we found that  $\a$ is $0.99 \pm 0.02$ (WMAP7+BAO+$H_0$) and $0.99 \pm 0.04$ (WMAP7). These values do not exclude possibility of acceleration.
Finding value of $\a$ is neither dependent of the background dynamics nor the dark energy models, therefore, in general, the power-law cosmology is not ruled out at late time. Larger number of SNIa data points (e.g. Union2) should be used in analysis with $H(z)$ data so that  one can tell more precisely whether the open power-law cosmology is favored over the flat and closed ones. Moreover, SNIa combined analysis with WMAP is recommended for identifying the cosmic geometry of the power-law cosmology.  When considering specific model of
scalar-field power-law cosmology in which canonical (quintessential) field evolving under potential and a dust fluid are major ingredients, we find field potential and the field velocity. These enable us to predict 
present value of $w_{\phi, 0}$ using CMB derived data in scenario of the scalar-field power-law cosmology. The predictions are $w_{\phi, 0} = -0.4489\pm0.0172 $  (WMAP7+BAO+$H_0$)
and $w_{\phi, 0} =  -0.4493 \pm 0.0300 $. (WMAP7). 
These results do not match model-independent WMAP7 $w$CDM  results (spatially flat) which are $w_{\phi, 0} = -1.12^{+0.42}_{-0.43}$  (WMAP7, constant $w$), $w_{\phi, 0} = -1.10^{+0.14}_{-0.14} $ (68 \% CL) (WMAP7+BAO+$H_0$, constant $w$).  We see that in $w$CDM model (constant $w$), the maximum observational allowance are $w_{\phi, 0} = -0.96$ (WMAP7+BAO+$H_0$) and $w_{\phi, 0} = -0.70$ (WMAP7) which are off the power-law cosmology's prediction. However in case of varying equation of state ($w=w(a)$), the combined CMB result gives $w_{\phi, 0} = -1.34^{+1.74}_{-0.36}$ (WMAP7+BAO+$H_0$+SN, ) and  $w_{\phi, 0} = -1.31^{+1.67}_{-0.38}$ (WMAP7+BAO+$H_0$+SN with time delay distance information) which allow the power-law cosmology within 1$\sigma$ uncertainty. It should be noted that, based on the $\Lambda$CDM model, the recent Planck collaboration result (Planck+WMAP polarization at low multipoles with 68\% CL) \cite{Ade:2013zuv} gives less value of present expansion rate, i.e. $H_0 = 67.3\pm 1.2$  km/sec/Mpc.  
The other cosmological parameters are $t_0 = 13.817\pm 0.048$, $\Omega_{\rm m, 0} = 0.315^{+0.016}_{-0.018}$. These parameters give approximately $\rho_{\rm c, 0} \approx 8.51\times 10^{-27}$ kg$/{\rm m}^3$, $\rho_{\rm m, 0} \approx 2.68 \times 10^{-27}$ kg$/{\rm m}^3$, $\alpha \approx 0.950$ and $w_{\phi,0} \approx -0.436$ (power-law cosmology prediction).    
The less $H_0$ affects the exponent $\alpha$ to be less. With Planck data, similar further work could also be done to test the quintessential power-law cosmology.

\section*{Acknowledgments}
The author thanks Emmanuel N. Saridakis for discussion, Chakkrit Kaeonikhom for assisted graphic works and the referee for useful comments. He thanks Reza Tavakol for hospitality at Astronomy Unit, Queen Mary University of London. The author is sponsored under project number: BRG5380018 of the Basic Research Grant Scheme of the Thailand Research Fund.

\appendix

\section{Observational data and constraints}
\label{Observational data and constraints}

A review the main sources of
observational constraints used in this work, WMAP7 Cosmic
Microwave Background (CMB), Baryon Acoustic Oscillations (BAO),
and Observational Hubble Data ($H_0$) is given here. In our calculations we take
the total likelihood $L\propto e^{-\chi^2/2}$ to be the product of
the separate likelihoods of BAO, CMB and $H_0$. Thus, the total
$\chi^2$ is
\begin{eqnarray}
\chi^2(p_s)=\chi^2_{\rm CMB}+\chi^2_{\rm BAO}+\chi^2_{H_0}.
\end{eqnarray}
\\

{\it{a. CMB constraints}}\\

We use the CMB data to impose constraints on the parameter space,
following the recipe described in Ref. \cite{Komatsu:2008hk2}. 
 The ``CMB
shift parameters'' \cite{Wang1} are defined as:
 \be
  R\equiv
\sqrt{\Omega_{\rm m, 0}}H_0 r\(z_*\),\,\quad l_{a}\equiv \pi
r\(z_*\)/r_{s}\(z_*\).
 \ee
  $R$ can be physically interpreted as a
scaled distance to recombination, and $l_{a}$ can be interpreted
as the angular scale of the sound horizon at recombination. $r(z)$
is the comoving distance to redshift $z$ defined as
 \be
r(z)\equiv\int_{0}^{z}\frac{1}{H\(z\)} {\rm d} z,
 \ee
 while $r_{s}\(z_*\)$ is
the comoving sound horizon at decoupling (redshift $z_*$), given
by
 \be
r_{s}\(z_*\)=\int_{z_*}^{\infty}\frac{1}{H\(z\)\sqrt{3\(1+R_{b}/\(1+z\)
\)}}  {\rm d} z.
 \ee
  The quantity $R_b$ is the ratio of the energy density
of photons to baryons, and its value can be calculated as
$R_b=31500 \Omega_{b,0} h^2 \(T_{\rm CMB}/2.7 \; {\rm K}\)^{-4}$, ($\Omega_{b, 0}$
being the present day density parameter for baryons) using
$T_{\rm CMB}=2.725$ (Refs. \cite{Komatsu:2008hk} and \cite{Komatsu:2008hk2}). The redshift at decoupling
$z_*\(\Omega_{b,0},\Omega_{\rm m,0},h\)$ can be calculated from the
following fitting formula \cite{husugiyama}:
 \be
z_*=1048\[1+0.00124\(\Omega_{b,0}
h^2\)^{-0.738}\]\[1+g_1\(\Omega_{\rm m, 0} h^2\)^{g_2}\], \ee with $g_1$
and $g_2$ given by:
\begin{eqnarray*}
g_1&=&\frac{0.0783\(\Omega_{b,0} h^2\)^{-0.238}}{1+39.5\(\Omega_{b,0} h^2\)^{0.763}}\\
g_2&=&\frac{0.560}{1+21.1\(\Omega_{b,0} h^2\)^{1.81}}.
\end{eqnarray*}
Finally, the $\chi^2$ contribution of the CMB reads
 \be
\chi^{2}_{\rm CMB}=\mathbf{V}_{\rm CMB}^{\mathbf{T}}\mathbf{C}_{\rm
inv}\mathbf{V}_{\rm CMB}. \ee
 Here $\mathbf{V}_{\rm
CMB}\equiv\mathbf{P}-\mathbf{P}_{\rm data}$, where $\mathbf{P}$ is
the vector $\(l_{a},R,z_{*}\)$ and the vector $\mathbf{P}_{\rm
data}$ is formed from the WMAP $5$-year maximum likelihood values
of these quantities \cite{Komatsu:2008hk} \cite{Komatsu:2008hk2}. The inverse covariance
matrix $\mathbf{C}_{\rm inv}$ is also provided in Refs.
\cite{Komatsu:2008hk} and \cite{Komatsu:2008hk2}.
\\

{\it{b. Baryon Acoustic Oscillations constraints}}\\

In this case the measured quantity is the ratio
$d_z=r_{s}\(z_{d}\)/D_{V}\(z\)$, where $D_{V}\(z\)$ is the so
called ``volume distance'', defined in terms of the angular
diameter distance $D_{A}\equiv r\(z\) /\(1+z\)$ as \be
D_{v}\(z\)\equiv\left[\frac{\(1+z\)^2 D_{A}^{2}(z) z
}{H(z)}\right]^{1/3},
 \ee
and $z_d$ is the redshift of the baryon drag epoch, which can be
calculated from the fitting formula \cite{HuEisenstein}: \be
z_d=\frac{1291\(\Omega_{\rm m, 0} h^2\)^{0.251}}{1+\(\Omega_{\rm m,0}
h^2\)^{0.828}}\[1+b_1\(\Omega_{b,0} h^2\)^{b_2}\],
 \ee
where $b_1$ and $b_2$ are given by
\begin{eqnarray*}
b_1&=&0.313\(\Omega_{\rm m, 0} h^2\)^{-0.419}\[1+0.607\(\Omega_{\rm m, 0} h^2\)^{0.674}\]\\
b_2&=&0.238\(\Omega_{\rm m, 0} h^2\)^{0.223}.
\end{eqnarray*}

We use the two measurements\cite{Percival:2009xn} of $d_z$ at redshifts $z=0.2$ and
$z=0.35$. We calculate the $\chi^2$
contribution of the BAO measurements as: \be
\chi^{2}_{\rm BAO}=\mathbf{V}_{\rm BAO}^{\mathbf{T}}\mathbf{C}_{\rm
inv}\mathbf{V}_{\rm BAO}.
 \ee
    Here the vector $\mathbf{V}_{\rm
BAO}\equiv\mathbf{P}-\mathbf{P}_{\rm data}$, with
$\mathbf{P}\equiv \( d_{0.2},d_{0.35} \) $, and $\mathbf{P}_{\rm
data}\equiv\(0.1905, 0.1097\)$, the two measured BAO data points
\cite{Percival:2009xn}. The inverse covariance matrix is provided
in Ref. \cite{Percival:2009xn}.\\

{\it{c. Observational Hubble Data constraints}}\\

The observational Hubble data are based on differential ages of
the galaxies \cite{ref:JL2002}. In Ref. \cite{ref:JVS2003}, Jimenez
{\it et al.} obtained an independent estimate for the Hubble
parameter using the method developed in Ref. \cite{ref:JL2002}, and
used it to constrain the equation of state of dark energy. The
Hubble parameter, depending on the differential ages as a function
of the redshift $z$, can be written as
\begin{equation}
H(z)=-\frac{1}{1+z}\frac{ {\rm d}  z}{ {\rm d}  t}.
\end{equation}
Therefore, once ${\rm d}z/{\rm d}t$ is known, $H(z)$ is directly obtained
\cite{ref:SVJ2005}. By using the differential ages of
passively-evolving galaxies from the Gemini Deep Deep Survey
(GDDS) \cite{ref:GDDS} and archival data \cite{ref:archive1} \cite{ref:archive2} \cite{ref:archive3} \cite{ref:archive4} \cite{ref:archive5} 
Simon {\it et al.}  \cite{ref:SVJ2005} obtained $H(z)$ in the range of $0\lesssim z
\lesssim 1.8$. We use the twelve observational
Hubble data from Refs. \cite{ref:0905} and \cite{Stern}  listed in Table
\ref{Hubbledata}.

\begin{table*}[!h]
\begin{ruledtabular}\begin{tabular}{ccccccccccccc}
 $z$ &\ 0 & 0.1 & 0.17 & 0.27 & 0.4 & 0.48 & 0.88 & 0.9 & 1.30 & 1.43 & 1.53 & 1.75  \\ \hline
 $H(z)\ ({\rm km~s^{-1}\,Mpc^{-1})}$ &\ 74.2 & 69 & 83 & 77 & 95 & 97 & 90 & 117 & 168 & 177 & 140 & 202  \\ \hline
 $1 \sigma$ uncertainty &\ $\pm 3.6$ & $\pm 12$ & $\pm 8$ & $\pm 14$ & $\pm 17$ & $\pm 60$ & $\pm 40$
 & $\pm 23$ & $\pm 17$ & $\pm 18$ & $\pm 14$ & $\pm 40$ \\
\end{tabular}
\label{Hubbledata}
\caption{The observational $H(z)$
data~\cite{ref:0905} \cite{Stern}.} \end{ruledtabular}
\end{table*}

The best-fit values of the model parameters from
observational Hubble data \cite{ref:SVJ2005} are determined by
minimizing
\begin{equation}
\chi_{H_0}^2(p_s)=\sum_{i=1}^{12} \frac{[H_{\rm th}(p_s;z_i)-H_{
\rm obs}(z_i)]^2}{\sigma^2(z_i)},\label{eq:chi2H}
\end{equation}
where $p_s$ denotes the parameters contained in the model,
$H_{\rm th}$ is the predicted value for the Hubble parameter,
$H_{\rm obs}$ is the observed value, $\sigma(z_i)$ is the standard
deviation measurement uncertainty, and the summation runs over the
$12$ observational Hubble data points at redshifts $z_i$.

\end{document}